\documentstyle[12pt]{report}
\begin{document}
\newcommand{\beq}{\begin{equation}}
\newcommand{\eeq}{\end{equation}}
\newcommand{\dsp}{\displaystyle}
\newcommand{\ap}{\alpha}

\vspace{1cm}

\hfill DSF-T-92/25 \\
\hfill INFN-NA-IV-92/25

\begin{center}
\vskip 3cm
{ \LARGE \bf Physical nonlinear aspects of classical \\
and quantum q-oscillators}
\vskip 1.8cm
{\Large V.I.Man'ko \footnote{on leave from Lebedev Physical Institute, Moscow},
G.Marmo, S.Solimeno, F.Zaccaria\\ Dipartimento di Scienze Fisiche\\
Univ. di Napoli Federico II \\ and I.N.F.N., Sez. di Napoli}\\
\vskip 0.5cm
\end{center}

\setcounter{footnote}{1}

\begin{abstract}
The classical limit of quantum q-oscillators suggests an interpretation
of the deformation as a way to introduce non linearity. Guided by this
idea, we considered q-fields, the partition fumction, and compute a
consequence on specific heat and second order correlation function
of the q-oscillator which may serve for
 experimental checks for the non linearity.
\end{abstract}

\newpage

\noindent{\bf 1. Introduction}
\setcounter{chapter}{1}
\bigskip

Recently the concept of quantum groups was developed in connection with the
solution of nonlinear equations [1-5]. The special case of the quantum
Heisenberg-Weyl group which produced the notion of quantum q-oscillator was
studied by Biedenharn [6] and Macfarlane [7]. They introduced q-oscillators
starting from generalized commutation relations containing an extra
dimensionless  parameter q. The physical sense of the quantum q-oscillator
up to now was not clarified. The attempt to connect quantum q-oscillators
with relativistic oscillator model has been discussed in [8]. The attempt
to introduce q-oscillators and quantum group SU(2) in the frame of the
generalized James-Cummings model has been discussed in [9]. Neverthless
these attempts are based on pure mathematical properties of q-oscillators.
The aim of our work is to show that the q-oscillator is a special case of
a nonlinear quantum oscillator.

The nature of q-oscillators of the electromagnetic field is clarified by the
nonlinearity of field vibrations.  The partition function for both classical
and quantum q-oscillator is calculated.

The spirit of the electromagnetic field q-quantum oscillator reminds us of
such a model of electrodynamics as the nonlinear Born-Infeld model [10].
In addition to clarify the physical nature of quantum q-oscillator as the usual
quantum oscillator with a "nonlinearity" of special type, the aim of this work
is to study the classical q-oscillator. This turns out to be the classical
nonlinear oscillator with the frequency depending on the amplitude by the
particular functional dependence. We will show that in this sense the
classical q-oscillator acquires a dependence on the amplitude of
 the physical pendulum. Assuming that the electromagnetic
 field be described by the
q-oscillator (both -quantum and classical ones) we will study the physical
consequences of this suggestion, namely how the parameters of the
q-oscillators change the  specific heat formula. This formula
reflects the linear properties of  the vibration of the vacuum at small
amplitudes. It is clear that at very  high intensities there must be
influence of the nonlinearity of the vacuum vibrations. We will calculate the
corrections to the partition function and to the specific heat formula.\\
As for notations, we shall denote by $\hbar$ the deforming parameter, to
avoid introducing a new symbol with respect to those already widely used
in the literature [1-7]. We apologize with reader for this. We ask him
to keep in mind  that Planck's constant is set equal to 1 everywhere in
paper.

\bigskip

\bigskip

\setcounter{chapter}{2}
\setcounter{equation}{0}
\noindent{\bf 2. Quantum q-oscillator}

\bigskip

\noindent
Let us introduce the usual creation and annihilation oscillator operators $a$
and $a^{\dag}$ obeying bosonic commutation relations
\begin{equation}
[a,a^{\dag}]=1 ~.
\end{equation}
The classical dynamical variables to which
$a$  and $a^{\dag}$  correspond oscillate with a frequency $ \omega = 1 $.
It is known that the operators $a,a^{\dag},1$ form the Lie algebra of
Heisenberg-Weyl group. So, the linear harmonic oscillator  may be connected
with the generators of pure Heisenberg-Weyl Lie group. In view of the
commutation relation (2.1) the usual scheme for generating the states of the
harmonic  oscillator is based on the properties of the Hermitean number
operator  $\hat{n}=a^{\dag}a$
\begin{equation}
[a,\hat{n}]=a,~~~[a^{\dag},\hat{n}]=-a^{\dag} ~.
\end{equation}
Thus
constructing the vacuum state $|0 \rangle$ obeying the equation
\begin{equation}
a|0 \rangle =0
\end{equation}
and the excited states
\begin{equation}
|n \rangle =\frac{a^{\dag~n}} {\sqrt{n!}}|0 \rangle
\end{equation}
which are eigenstates of the number operator $\hat{n}$
\begin{equation}
\hat{n}|n \rangle =n|n \rangle, n \in Z^{+}
\end{equation}
the matrix representation of the operators a and $a^{\dag}$ in the basis (2.4)
have  the known
expressions
\begin{eqnarray}
a&=&\left(\begin{array}{crcl}
0 & \sqrt{1} & 0 & \ldots \\
0 & 0 & \sqrt{2} & 0 \\
0 & 0 & 0 & \sqrt{3} \\
\ldots & \ldots & \ldots & \dots
\end{array}\right)\nonumber\\
a^{\dag}&=&\left(\begin{array}{crcl}
0 & 0 & 0 & \ldots \\
\sqrt{1} & 0 & 0 & \ldots \\
0 &\sqrt{2} & 0 & \ldots \\
\ldots & \ldots & \ldots & \ldots
\end{array}
\right)
\end{eqnarray}
while the number operator $\hat{n}$ is described by the matrix
\begin{equation}
\hat{n}=\left(\begin{array}{crcl}
0 & 0 & 0 & \ldots \\
0 & 1 & 0 & \ldots \\
0 & 0 & 2 & \ldots \\
  . & . & . & \ldots \\
\end{array}\right)~~~.
\end{equation}
The Hamiltonian for such a system is defined as
\begin{equation}
H = \omega \, \frac{a^{\dagger}a + aa^{\dagger}}{2} ~~~~~ \omega
> 0~~~~.
\end{equation}
The q-oscillators may be introduced by generalizing
 the matrices (2.6) and (2.7)
with the help of the q-integer numbers $n_{q}$,
\begin{equation}
n_{q}=\frac{\sinh~n\hbar}{\sinh~\hbar},~~~q=e^{\hbar} ~~~.
\end{equation}
Here $\hbar$ and $q$ are dimensionless $c$-numbers,
 which appear at this
 purely mathematical level. When $\hbar=0$, q=1 and the q-integer
$n_{q}$ coincides with $n$. Then, replacing the integers in (2.6) and (2.7)
by q-integers we obtain  matrices which define  the annihilation and
creation operators of the quantum q-oscillator,
\begin{eqnarray}
a_{q}&=&\left(\begin{array}{crcl}
0 &  \sqrt{1_{q}} & 0 & ...\\
0 & 0 & \sqrt{2_{q}} &...\\
0 & 0 & 0& \sqrt{3_{q}}\\
   ...&...&...&...
\end{array}\right)\nonumber\\
a^{\dag}_{q}&=&\left(\begin{array}{crcl}
0&0&0&...\\
\sqrt{1_{q}}&0&0&...\\
0&\sqrt{2_{q}}&0&... \\
   ...&...&...&...
\end{array}\right)\nonumber\\
\hat{n}_{q}&=&\left(\begin{array}{crcl}
0&0&0&...\\
0&1_{q}&0&...\\
0&0&2_{q}&...\\
   ...&...&...&...
\end{array}\right)~~.
\end{eqnarray}
since the action of $\hat n_q$ on eigenstates $|n>$ is given by
\beq
\hat n_q |n> = \frac {\sinh n\hbar}{\sinh \hbar}|n>\nonumber\\
\eeq
The above matrices obey the commutation relation
\begin{equation}
[a_{q},\hat{n}]=a_{q},~~~[a_{q}^{\dag},\hat{n}]=-a_{q}^{\dag}
\end{equation}
but the commutation relations of the operators $a_{q}$ and $a_{q}^{\dag}$ do
not
coincide with the boson commutation relations. Equation (2.1) is replaced by
\begin{equation}
[a_{q},a_{q}^{\dag}] = F(\hat{n})
\end{equation}
where the function $F(\hat{n})$ has the form
\begin{equation}
F(\hat{n}) = f^{2}(\hat{n}+1)- f^{2}(\hat{n})
\end{equation}
For $\hbar=0$ (2.12) reduces to (2.1).\\ In addition
to the above commutation relation there exists the reordering relation
\begin{equation}
a_{q}a_{q}^{\dag}-qa_{q}^{\dag}a_{q}=q^{-\hat{n}}
\end{equation}
which usually is taken as the definition of q-oscillators.

It is
worthy noting that the operators $a_{q}$ and $a_{q}^{\dag}$
can be expressed in terms of the operators $a$ and $a^{\dag}$ [11-13]
\begin{equation}
a_{q}= af(\hat{n}),~~~a_{q}^{\dag}=f(\hat{n})a^{\dag},
\end{equation}
where (see Eq. (2.9))
\begin{equation}
f(\hat{n})=\sqrt{\frac{\hat{n}_q}{\hat{n}}}
\end{equation}
We have also
\begin{equation}
\hat{n}_{q} = a_{q}^{\dag}a_{q}
\end{equation}
and
\begin{equation}
[a_{q},\hat{n}_{q}] = F(\hat{n}) a_{q}
{}~~~~~~~~~~~~~~[ a_{q}^{\dag}, \hat{n}_q] = - a_{q}^{\dag} F(\hat{n})
\end{equation}

 In the Schr\"{o}dinger representation the evolution operator of the harmonic
oscillator
\begin{equation}
U(t)=\exp \left[ -i\omega\, \frac{(a^{\dag}a + aa^{\dag})t}{2} \right]
\end{equation}
gives the possibility to find out explicitly linear integrals of motion which
depend on time [14,15],
\begin{eqnarray}
A(t)&=&U(t)aU^{-1}(t)=e^{i \omega t}a\nonumber\\
A^{\dag}(t)&=&U(t)a^{\dag}U^{-1}(t)=e^{-i \omega t}a^{\dag}~~~.
\end{eqnarray}
The matrices of the integrals of motion (2.21) in Fock basis may be
obtained from the equations
\begin{equation}
\\A(t)|n\rangle = e^{i \omega t} \sqrt{n} |n-1\rangle,
A^{\dag}(t) |n\rangle = e^{-i \omega t} \sqrt {n+1} |n+1\rangle
\\
\\
\end{equation}
Let us now introduce the Hamiltonian
\begin{equation}
\hat{H}=\omega \frac{a_{q}a_{q}^{\dag} + a_{q}^{\dag}a_{q}}{2}
\end{equation}
for which the evolution operator takes the form
\begin{equation}
U_{q}(t)=\exp \left[ -i\omega t\frac{(a_{q} a_{q}^{\dag} +
a_{q}^{\dag}a_{q})}{2} \right]~.
\end{equation}
We have for the integrals of motion
\begin{equation}
A_{q}(t)=U_{q}a_{q}U_{q}^{-1},  ~~~~A^{\dag}=U_{q}a_{q}^{\dag}U_{q}^{-1}
\end{equation}
the following matrix expressions,
\begin{eqnarray}
A_{q}(t)&=&\left(\begin{array}{cccc}
0 & ~~\sqrt{1_{q}}e^{i(1_{q}-0_{q})\omega t}~~ & 0 &  \ldots \\
0 & 0 & ~~\sqrt{2_{q}}e^{i(2_{q}-1_{q})\omega t}~~ & \ldots \\
{}~~\ldots~~ & \ldots & \ldots & \ldots\\
\end{array}\right)\nonumber\\
A_{q}^{\dag}(t)&=&\left(\begin{array}{cccc}
0 & 0 & 0 & \ldots\\
\sqrt{1_{q}}e^{-i(1_{q}-0_{q})\omega t}~~  & 0 & 0 & \ldots\\
0 & \sqrt{2_{q}}e^{-i(2_{q}-1_{q})\omega t}~~& 0 & \ldots\\
{}~~\ldots~~ & \dots & \dots & \ldots\\
\end{array}\right) ~~~.\nonumber\\
\end{eqnarray}
These operators are the generalizations of the linear integrals of motion
(2.21) to the case of nonlinear Hamiltonian.

In general we can prove the following string of relations
\begin{eqnarray}
U^{-1}a_{q}&=&a_{q}U^{-1}_{-1}\nonumber\\
U^{-1}a^{\dag}_{q}&=&a^{\dag}_{q}U^{-1}_{1}\nonumber\\
Ua_{q}&=&a_{q}U_{-1}\nonumber\\
Ua^{\dag}_{q}&=&a^{\dag}_{q}U_{1}
\end{eqnarray}
where $U_{\pm p}$ is defined implicitely by expressing $U$ as a function of
the operator $\hat{n}$,
\begin{equation}
U_{\pm p}(\hat{n})\equiv U(\hat{n}\pm p)
\end{equation}
The integrals of motion $A_{q}(t),A_{q}^{\dag}(t)$, which coincide with the
operators $a_{q},a^{\dag}_{q}$ at $t=0$, may be expressed in terms of the
matrices (2.6) and (2.7) by means of the relations (2.26) and (2.27).

\setcounter{chapter}{3}
\setcounter{equation}{0}

\bigskip

\bigskip

\noindent {\bf 3. Classical q-variables}

\bigskip

\noindent
In this section we will use the inverse of the procedure proposed by Dirac
to quantize a classical system, i.e. we consider a "dequantization
procedure". We will obtain therefore a dynamical
system on the classical phase space $(q,p)$, or $(\alpha, \alpha^*)$,
from the quantum system of the previous section, the latter then being
the Dirac quantized version of the former.

The usual harmonic oscillator vibrating with unit frequency
may be described in terms of the following variables
\begin{equation}
\begin{array}{l}
{\displaystyle \alpha = \frac{ q + ip}{\sqrt{2}} } \\
{}~~ \\
{\displaystyle \alpha^* =\frac{q - ip}{\sqrt{2}}~,}
\end{array}
\end{equation}
with non-zero Poisson bracket
\begin{equation}
\{\alpha, \,\alpha^*\} = -i
\end{equation}
and Hamiltonian function
\begin{equation}
H = \alpha \alpha^*~.
\end{equation}
The flow associated with equations of motion is
\begin{equation}
\alpha = A e^{i (t + \varphi)} ~~~~~ \alpha^* = A e^{-i (t + \varphi)}
\end{equation}
where
\begin{eqnarray}
& & A = \alpha (0) \alpha (0)^* \nonumber \\
& & \varphi = \frac{1}{2i} \, \ln \, \frac{\alpha
(0)}{\alpha (0)^*}
\end{eqnarray}

For a frequency different from 1, the Hamiltonian is
\beq
H = \frac{p^2 + \omega^2 q^2}{2} ~~~~~~~~~~~~\omega >0
\eeq
In terms of complex variables, we  have now
\begin{eqnarray}
& & \ap = \left( \frac{ip}{\sqrt{\omega}} +  \sqrt{\omega} q
\right) \frac{1}{\sqrt{2}} ~~~~~~~~\ap^* = \left( \frac{-ip}{\sqrt{\omega}}
+  \sqrt{\omega} q \right) \, \frac{1}{\sqrt{2}} \nonumber \\
& & H(\ap, \ap^*) = \omega \ap \, \ap^* \nonumber \\
& & \{\ap, \, \ap^*\} = -i
\end{eqnarray}
{}From (2.16) we define the classical q-oscillator in terms of these new
variables $(\ap, \, \ap^*)$, namely
\beq
\ap_q = \sqrt{\frac{\sinh \hbar \ap \ap^*}{\ap \ap^* \, \sinh \hbar}} \ap
{}~~~~~~~\ap^*_q = \sqrt{\frac{\sinh \, \hbar \ap \ap^*}{\ap \ap^* \, \sinh
\hbar}} \ap^*
\eeq
Their Poisson bracket can be computed and expressed in terms of themselves,
obtaining
\beq
\{ \ap_q, \, \ap^*_q\} = -i\frac { \hbar}{\sinh \hbar}\sqrt{1 +
 |\ap_q|^{4}(\sinh \hbar)^2}
\eeq
so that we will consider a new system described by such q-variables with
Hamiltonian function
\beq
H(\ap_{q},\ap^*_{q}) = \omega \ap_{q}\ap_{q}^*.
\eeq
The equations of motion are then
\beq
\dot \ap_q = - i \frac {\hbar}{\sinh \hbar} \omega \sqrt {1 + |\ap_{q}|^4
(\sinh \hbar)^2}\ap_q
\eeq
and complex conjugated, with solutions

\beq
\alpha_{q}(t) = \alpha_{q}(0) exp [\frac {-it \omega \hbar}{\sinh \hbar}
\sqrt {1 + \mid \alpha_{q}(0)\mid^{4}( \sinh \hbar)^2}]
\eeq
and its conjugate.\\
All this can be commented stating that we have performed
a non canonical transformation, i.e. deformed the Poisson
bracket structure, while preserving the form
of the Hamiltonian function and note
that the above Poisson bracket has the standard bracket as
a limit when $\hbar$ goes to $0$.\\
Such a system can be, of course, re-written in terms of $(\ap, \ap^*)$
variables: in these coordinates the original Poisson bracket is unchanged
while the Hamiltonian function is
\beq
H_{q}(\ap, \ap^*) = \omega \frac {\sinh \hbar \ap \ap^*}{\sinh \hbar}
\eeq
This new dynamical system has a phase portrait which is the same
as the usual linear  harmonic oscillator. In facts, the
 equations of motion for the latter
system, which evolves under the Hamiltonian $H$ are
\beq
\dot \alpha =-i\omega\alpha~~~~~~
{\dot \alpha}^{*}=i\omega\alpha^*
\eeq
and for the system with the Hamiltonian $H_q$ are
\beq
\dot\alpha=-i \omega_q \alpha~~~~~~
{\dot \alpha}^* = i\omega_q \alpha^*
\eeq
with
\beq
\omega_q  =\omega \frac{\hbar}{\sinh \hbar} \, \cosh \hbar \ap \ap^* ~.
\eeq
We notice that $\alpha \alpha^*$ is a constant of the
motion for both systems. The main difference between such systems is that
the frequency for the second one depends on the orbit while for the
the first one is constant.\\
This leads to interpreting q-oscillators as systems carrying a particular
non-linearity. What in the harmonic oscillator is a constant $\omega$
characterizing the evolution along any orbit, becomes here a function
constant on each orbit separately, i.e.\ a constant of the motion for
the undeformed
harmonic oscillator. From the formulae (2.16) it follows a
possible  physical interpretation of the q-oscillator.
 The amplitude of the vibrations of the harmonic
oscillator is not connected with the phase of the vibrations which changes
linearly in time. For the q-oscillator the frequency depends
 on its amplitude. The formula (2.16) reflects this classical
nonlinear phenomenon for the quantum oscillator. Such a situation can be
generalized to any linear system.\\ We digress briefly to describe on what
would be the situation in general.

Let us have, indeed, a linear system
\beq
\dot{\underline{X}} = A \underline{X}
\eeq
where $\underline{X} \in {\bf R}^n$ and $A$ a $n \times n$ matrix with
constant entries. It can be integrated
\beq
\underline{X} (t) = e^{tA} \underline{X} (0)~.
\eeq
Its integrals of the motion can be expressed in terms of the initial
conditions $\underline{X} (0) = \underline{X}_{0}$. Replacing $A$ with
$B(\underline{X})$ defined by
\beq
[B(\underline{X})]^{i}_{j} = A^{i}_{j}f^{i}_{j}~~,
\eeq
where $f^{i}_{j}$ is any constant of the motion for the original linear
system, the new system, which is non linear,
\beq
\underline{\dot X} = B(\underline{X}) \underline{X}
\eeq
can be integrated via exponentiation, because for any
 initial condition $B(\underline{X}(0))$ is a matrix  constant along the
trajectory originated from it.\\
When the starting linear system is Hamiltonian, it turns out to be
completely integrable. Therefore, eq. (3.20) is a generalization
of our q-deformed oscillator. By employing various constants of the
motion we could accommodate deformations with many parameters.
There are other ways to get non linear systems out of linear ones, for
instance by  using a generalized reduction procedure [16]. In the future
we shall analyze and compare among them these procedures that use
deformation or reduction to get non linear systems.

\setcounter{chapter}{4}
\setcounter{equation}{0}

\bigskip
\bigskip

\noindent{\bf 4. Two dimensional q-oscillators}
\bigskip
\noindent

In this section we will analyze a possible q-deformation of the two
 dimensional oscillator. There exists
two essentially different possibilities to introduce the q-deformed creation
and annihilation operators.They are based on our interpretation of
 the q-oscillators
as describing nonlinear vibrations. If two modes in the linear limit
 have different
frequences it is natural to deform the annihilation and creation operators for
these two modes independently of each other, by simply using
 the formulae for
one-dimensional q-oscillator discussed above. But if we want to take into
account the isotropy of two dimensional oscillator and to conserve
 the $U(2)$ dynamical
symmetry even in the nonlinear vibration regime we might deform the two
dimensional oscillator in such a way that  both constituent oscillators
have the same frequency in the equal initial conditions for this regime too.
Taking these remarks into account let us introduce two sets of operators
\beq
a_{q \pm} = a_{\pm} f(\hat{n}), ~~
a^{\dagger}_{q \pm} = f(\hat{n})a^{\dagger}_{q \pm}
\eeq
Here the number operator $\hat{n}$ is defined as follows
\beq
\hat{n} = n_{+} + n_{-}
\eeq
where
\beq
n_{\pm} = a^{\dagger}_{\pm}a_{\pm}~
\eeq
and the operators $ a_{\pm},a^{\dagger}_{\pm} $ obey the usual boson
 commutation relations
\beq
[a_{\pm}, a^{\dagger}_{\pm}] = 1
\eeq
all the others being zero.
The commutation relations for the two components of the two-dimensional
 q-oscillator are then as follows
\beq
[a_{q+}, a_{q-}] = 0, ~~~~
[a_{q+}, a^{\dagger}_{q-}] = a_{+}a_{-}^{\dagger}[f^{2}(\hat{n}+1) - f^{2}(
\hat{n})]
\eeq

The other commutation relations are obtained from the conjugation of
the above.
The non-zero commutation relation of the annihilation and
 creation operators describing
the same component is of the form
\beq
[a_{q \pm}, a^{\dagger}_{q \pm}] = a^{\dagger}_{\pm}a_{\pm} (f^{2}(\hat{n}+1)
- f^{2}(\hat{n})) +
f^{2}(\hat{n}+1).
\eeq
Due to the nonlinear interaction of vibrations the operators $a_{q+}$ and $
a_{q-}
^{\dagger}$ do not commute, as well as
$a_{q-}$ and $a_{q+}^{\dagger}$. In the limit, when $\hbar$
 goes to zero (q goes to 1)
the written commutation relations lead to the usual boson commutation relations
since in this limit the function $f$ goes to 1.The four operators
\beq
    a_{+}^{\dagger}a_{+}, a_{-}^{\dagger}a_{-}, a_{+}^{\dagger}a_{-},
a_{-}^{\dagger}a_{+}
\eeq
form the generators of the U(2) group representation.This representation is
realized in the space of states of two dimensional harmonic oscillator.We
deformed the operator $a_+$ and $a_-$ and they became the operators $a_{q+}$
 and $a_{q-}$.
But the deformation given by the formulae (4.1) is such that the operators
\beq
   a_{q+}^{\dagger}a_{q+}, a_{q-}^{\dagger}a_{q-}, a_{q-}^{\dagger}a_{q+},
a_{q-}^{\dagger}a_{q+}
\eeq
realize the same representation of the same undeformed
 group U(2) in the space of the same
two dimensional harmonic oscillator states.This happens because
 these new generators
differ from the former ones only by the coefficients which depend on the
Casimir operator of the rotation group. So, for any irreducible subspace these
coefficients are simply c-numbers and the operators (4.10) close
on the Lie algebra
of the group U(2) in these subspaces. It can be shown that the following
re-ordering relation holds
\beq
a_{q+}a^{\dagger}_{q+} + a_{q-}a^{\dagger}_{q-} - q (a^{\dagger}_{q+}a_{q+}
+ a^{\dagger}_{q-}a_{q-}) = q^{-\hat{n}}
\eeq
Had we q-deformed the two components of the oscillator
as independent ones, namely as
those appearing in the previous section 2, the right hand side of
the Eq.(4.11) would  be replaced by the term $q^{-\hat{n}_{+}} +
q^{-\hat{n}_{-}}$  . This is a different instance of the non
 linearity of the q-oscillators.\\ If we deal with many more oscillators,
we can use mixtures of these q-deformation procedures according to
which linear transformations we want to survive  after the
 deformation process.
Let us clarify this statement. We now study transformations
 between two-dimensional
q-oscillators which are analogous to canonical transformations of linear
oscillators, $(a_{\pm}, a_{\pm}^{\dagger})$ and $(b_{\pm}, b_{\pm}^{\dagger})$,
by requiring that
\beq
a_{+}a_{+}^{\dagger} + a_{-}a_{-}^{\dagger} = b_{+}b_{+}^{\dagger} +
b_{-}b_{-}^{\dagger} ~~.
\eeq
The resulting linear transformations are
\begin{equation}
a_{\pm}=A_{\pm}b_{+}+B_{\pm}b_{-}
\end{equation}
\begin{equation}
a^{\dagger}_{\pm} = A^*_{\pm} b_{+}^{\dagger} + B^*_{\pm} b^{\dagger}_{-}
\end{equation}
with $A_{\pm}$ and $B_{\pm}$ complex numbers such that the matrix

$$
\left(\matrix{      A_{+}&B_{+} \cr
                    A_{-}&B_{-} \cr}\right)
$$
belongs to U(2).
Substituting the above formulae in (4.l) and (4.2) we have consequently

\beq
a_{q \pm} = (A_{\pm}b_{+}+B_{\pm}b_{-})f(\hat{n})
\eeq
\beq
a^{\dagger}_{q \pm} =
f(\hat{n})(A_{\pm}^{*}b_{+}^{\dagger}+B_{\pm}^{*}b_{-}^{\dagger})
\eeq
Here the number operator $\hat{n}$ is written as follows
\beq
\hat{n} =\hat{n}_{b+} +\hat{n}_{b-}
\eeq
where
\beq
\hat{n}_{b+} = b^{\dagger}_{+}b_{+}
\eeq
\beq
\hat{n}_{b-} = b^{\dagger}_{-}b_{-}~~.
\eeq
If we consider a dynamical system described by the Hamiltonian,
\begin{equation}
H = \frac{1}{2}(a_{q+}^{\dagger}a_{q+}+a_{q-}^{\dagger}a_{q-}+
a_{q+}a_{q+}^{\dagger}+a_{q-}a_{q-}^{\dagger})~~.
\end{equation}
Written in terms of the number operator, $H$ acquires the following
form
\begin{equation}
H= \frac{1}{2}\left[\hat{n}f^{2}(\hat{n}-1)+(\hat{n}+2)f^{2}(\hat{n}+1)\right]
\end{equation}
which is manifestly invariant under the previous transformations.
In the Fock basis $|n_+>|n_->$ the eigenvalues of $H$ are immediately
computed as $\hat{n}$ has positive integers eigenvalues and each of them
has degeneracy $n+1$.\\
  It is obvious how to define the variables for
the classical 2-dimensional q-oscillators. For instance
\beq
\alpha_{q\pm} = \alpha_{\pm}f(n)
\eeq
where $(\alpha_{\pm}, \alpha_{\pm}^*)$ are variables for the usual
harmonic oscillators and
\beq
n = n_{+} + n_{-}
\eeq
with
\beq
n_{\pm} = \mid\alpha_{\pm}\mid^{2}~~.
\eeq
  The computation of the non-zero Poisson brackets then
gives the following result
\beq
\{\alpha_{q\pm}, \alpha^*_{q\mp}\} = -i \alpha_{\pm}\alpha^*_{\mp} \frac
{(\hbar n)\cosh n\hbar - \sinh n\hbar}{n^{2} \sinh \hbar}
\eeq
and
\beq
\{\alpha_{q\pm},\alpha^*_{q\pm}\} = \frac {-i}{n \sinh \hbar}[ (1 - \frac
{n_{\pm}}{n})\sinh n\hbar + \hbar n_{\pm} \cosh n\hbar]~~~.
 \eeq
Thus, it is clear that those transformations which preserve the term
responsible for the non linearity will be compatible with the q-deformation.
More explicitely, this means that these transformations commute with the
deformation procedure.

\setcounter{chapter}{5}
\setcounter{equation}{0}

\bigskip

\bigskip

\noindent{\bf 5. q-fields}
\bigskip

\noindent
We take into consideration the usual expansions of classical and
quantum fields on an interval of length $L=1$ in terms of usual harmonic
oscillators. For simplicity we start considering a classical real
 massless scalar field
$\phi (x,t)$ and its conjugate $\pi (x,t)$, satisfying also $\dot\phi =
\pi$
\beq
\phi (x,t) = \sum_j^{-\infty, +\infty} e^{ik_jx} q_j (t) ~~~~~~~~\pi (x,t)
= \sum_j^{-\infty, + \infty} \, e^{-ik_jx} p_j (t) ~~.
\eeq
Notice that here $q_{j}$ and $p_{j}$ are complex variables which anyhow
can be expressed in terms of harmonic oscillators as, for instance in
[17] so that also
in this case the usual Poisson bracket holds
\beq
\{q_i, p_j\} = \delta_{ij} ~~~~~~k_j = 2 \pi j~.  \eeq
The use of such fields is aimed at a study of some aspects of
 electrodynamics with massless
 q-photons.
Following [17], we have
\begin{eqnarray}
& & \phi (x, t) = \sum_j e^{ik_j x} \, \frac{1}{\sqrt{2 \omega_j}}
 (\ap^*_{-j} +
\ap_j) \nonumber \\
& & \pi (x,t) = -i \sum_j e^{-ik_j x} \sqrt{\frac{\omega_j}{2}} \,
(\ap_{-j} -\ap_{j}^*) \nonumber \\
& & \omega_j = |2 \pi j |~~  .
\end{eqnarray}
 These variables $(\alpha_j, \alpha^*_j)$ with $j=\pm$ integers can
be written in terms of usual oscillator variables for positive and
negative frequencies $(\alpha_{j\pm}, \alpha^*_{j\pm})$ with $j$
positive integers, namely
\beq
\alpha_j = \frac{\alpha_{j+} -i \alpha_{j-}}{\sqrt 2} ~~,
\eeq
\beq
\alpha_{-j} = \frac {\alpha_{j+} +i \alpha_{j-}}{\sqrt2}
\eeq
and their complex conjugates. The Poisson brackets of the usual
oscillators are of the form
\beq
\{\alpha_{j\pm}, \alpha_{j'\pm}\} = 0~~,
\eeq
\beq
\{\alpha_{j\pm}, \alpha^*_{j'\mp}\} = 0~~,
\eeq
and
\beq
\{\alpha_{j\pm}, \alpha^*_{j'\pm}\} = -i \delta_{j,j'}~~.
\eeq
 We have
\beq
\{ \phi (x,t), \, \pi (x', t) \} = \delta (x - x')~.
\eeq

We are now in a position to define a pair of classical q-fields,  noting
that there is not a unique definition due to the fact that the
frequency depends on $\ap_j \ap_j^*$.

By using previous equations, we write
\beq
\begin{array}{l}
{\dsp \phi_q (x,t) = \sum_j e^{i 2\pi j x} \, \frac{1}{\sqrt\omega_{jq}} \,
\frac{\ap_{-jq}^* + \ap_{jq}}{\sqrt{2}}} \\
{}~~ \\
{\dsp \pi_q (x,t) = -i \sum_j e^{-i 2 \pi jx} \,\sqrt \omega_{jq}
 \frac{-\ap_{jq}^* +
\ap_{-jq}}{\sqrt{2}} }
\end{array}
\eeq
where
\beq
\omega_{jq} = |2 \pi j| \, \hbar \, \frac{\cosh ( \ap_j
\ap_j^* + \ap_{-j}\ap^*_{-j})\hbar}{\sinh \hbar}
\eeq
and $\ap_{jq}$ and $\ap_{jq}^*$ are given as functions of $\ap_{j\pm}$ and
its complex conjugate; for instance,
\beq
\alpha_{jq} = \frac{\alpha_{j+} -i \alpha_{j-}}{\sqrt 2}\sqrt \frac
{\sinh (|\ap_{j+}|^2 +|\ap_{j-}|^2)\hbar}{(|\ap_{j+}|^2 +|\ap_{j-}|^2)
\sinh \hbar}~~~~.
\eeq
  The computation of the Poisson brackets for these fields gives
results which are different from the
usual ones, as it was expected, up to corrections of the order of $\hbar^
{2}$. The above fields have been constructed allowing a coupling
between positive and negative frequency modes belonging to the same
index $j$.\\ To give an idea of the effects of a q-deformation, we
give explicitely the behaviour of the Poisson bracket with $\hbar
\rightarrow 0$ between
conjugated fields, which have been q-deformed in
another way that allows for much simpler computations. Such new fields are
defined as follows
\beq
\phi_{q}(x,t) = \sum_j e^{i2\pi jx}(\frac{\alpha^*_{-jq}}{\sqrt {2
\omega_{-jq}}} + \frac {\alpha_{jq}}{\sqrt {2 \omega_{jq}}})
\eeq
\beq
\pi_{q}(x,t) = -i \sum_j e^{-i2\pi jx} ({\sqrt{\frac{\omega_{-jq}}{2}}}
\alpha_{-jq} - \sqrt {\frac {\omega_{jq}}{2}}\alpha^*_{jq})
\eeq
where
\beq
\omega_{jq} = |2\pi j| \frac{\hbar}{\sinh \hbar}\cosh (|\alpha_{j}|^{2}\hbar)
\eeq
and the q-deformed oscillators are each other independent and defined as
in (3.8). That is we are allowing only the self coupling.
We can compute the Poisson brackets between such fields at equal
times and obtain
\beq
\{\phi_{q}(x,t), \pi_{q} (x',t)\} = \delta (x-x')(1-\frac{\hbar^2}{6})
 +\hbar^{2}(c_{1} +c_{2})
\eeq
where
\beq
c_1 = \frac{1}{4} \sum_{j>0}[( e^{i2\pi j(x-x')}(|\alpha_{j}|^{4} +
|\alpha_{-j}|^{4})) + c.c.]
\eeq\
and
\beq
c_2 = \frac {1}{2} \sum_{j>0}[(e^{i2\pi j(x+x')}(\alpha_j^2|\alpha_{j}|^{4}
- \alpha_{-j}^{*2}|\alpha_{-j}|^{4})) + c.c.]~~.
\eeq

It can be shown that
 \beq
\{\phi_{q}(x,t), \phi_{q}(x',t)\} = O(\hbar^{2})
\eeq
\beq
\{\pi_{q}(x,t), \pi_{q}(x',t)\} = O(\hbar^{2})~~~.
\eeq
All this gives an account of the effect of q-nonlinearity as far as
the principle of locality is concerned. The corrections destroy the
$\delta$-function behaviour of the Poisson bracket between conjugated
fields and are proportional to the fourth power of the amplitudes of
nolinear modes vibrations. We shall come back to these aspects in the
future.
The quantized fields are defined from eqs. (5.13-14) replacing $c$-numbers
$(\ap_{jq}, \ap_{jq}^*)$  with the creation and annihilation operators
$(a_{jq}, a_{jq}^{\dagger})$ with $\omega_{jq}$ becoming also an
operator $\hat\omega_{jq}$,  namely
\beq
\hat{\phi}_q (x,t) = \sum_j e^{i2\pi jx}(\frac {1}{\sqrt{2\hat{\omega}_{jq}}}
a_{jq}+\frac{1}{\sqrt{2\hat{\omega}_{-jq}}}a^{\dagger}_{-jq})
\eeq
and
\beq
\hat{\pi}_q (x,t) = \sum_j e^{i2\pi jx}(\sqrt{\frac{\hat{\omega}_{-jq}
}{2}}a_{-jq} - \sqrt{\frac{\hat{\omega}_{jq}}{2}}a^{\dagger}_{jq})
\eeq
with
\beq
\hat{\omega}_{jq} = |2\pi j|\hbar \frac{\cosh (\frac{a_j^{\dagger}a_j + a_j
a_j^\dagger )}{2}\hbar)}{\sinh \hbar}
\eeq
Such a definition of quantized q-fields just follows the usual prescript-
ion, particularly for what concerns the $j=0$ term in the sum and holds
since the operators $\hat\omega_{jq}$ are invertible for $j\neq.0$.
 Their commutation
relations at equal time can be calculated and as it was easily expected,
we  have that such fields are not local.\\
It should be noticed that we have not addressed the problem of deformation
of the differential Maxwell equations, i.e. we are not exhibiting a
q-deformed d'Alembertian. As a matter of fact, at this stage we do not
even know whether our q-fields are solutions of some differential equation.
These and other related aspects will be taken up in a forthcoming paper.

\setcounter{chapter}{6}
\setcounter{equation}{0}

\bigskip

\bigskip

\noindent

\noindent {\bf 6. Partition and time-corrrelation functions}
\bigskip

\noindent
The partition function for a single q-oscillator
\begin{equation}
Q(\beta,\hbar)=\sum_{n=0}^{\infty} \exp(-\beta n_{q})
\end{equation}
is a function of the temperature, through the variable $\beta=\omega/kT$, and
the parameter
$\hbar$.
In particular, $\hbar$ is very small and the series representing Q can be
represented in analytic forms.
We will consider first the limiting expression of Q obtained by letting
$\hbar\rightarrow 0$ for fixed $\beta$ thus obtaining
a series representation of the form,
\begin{equation}
Q(\beta,\hbar)=Q_{0}+\frac{1}{2}\hbar^{2}\frac{d^{2}}{d\hbar^{2}} \, Q+...
\end{equation}
where $Q_{0}(\beta)=Q(\beta,0)=1/(1-e^{-\beta})$ represents the partition
function of the
ideal harmonic oscillator. In particular,
 we will see that the above series does not converge
for $\beta<\hbar$. In view of this, we will discuss the limit
 of $Q$ for $\hbar\rightarrow 0$ for given
value of the ratio $\zeta=\beta/\sinh~\hbar$. The quantum partition function
 obtained in this second case is well approximated by the classical one.

\bigskip

\subsection{\bf Partition function for given $\beta$}

\bigskip

When we keep $\beta$ fixed and let $\hbar$ go to zero we can represent $Q$
as a power series
in $\hbar$ by relying on the power expansion of $n_{q}$,
\begin{equation}
n_{q}=n+\frac{\hbar^{2}}{3!}(n^{3}-n)
+\frac{\hbar^{4}}{5!} \left( n^{5}-\frac{10}{3}n^{3}+\frac{7}{3}n
\right)+\cdots
\end{equation}
Then,
\begin{eqnarray}
Q(\beta,\hbar)&=&
\sum_{n=0}^{\infty} \exp(-\beta n_{q}+\beta n)
\exp(-\beta n)\nonumber\\
& \approx& Q_{0}(\beta)\left(1+\frac{W^{2}}{3!}\beta^{3}(Q_{3}-Q_{1})\right)
\nonumber\\
& = & \exp\left(\beta\frac{\sinh~(\hbar\partial/\partial\zeta)}{\sinh~\hbar}-
\beta\partial
/\partial\zeta\right)Q_{0}(\zeta)|_{\zeta=\beta}\nonumber\\
&\approx&Q_{0}(\beta)
\left(1+\frac{W^{2}}{3!}\beta^{3}(Q_{3}-Q_{1})\right.\nonumber\\
&+&\left.W^{4}\left(\frac{1}{3!^{2}}\beta^{6}(Q_{6}-2Q_{4}+Q_{2})
+\frac{1}{5!}\beta^{5}(Q_{5}-\frac{10}{3}Q_{3}+\frac{7}{3}Q_{1}) \right)\right)
\nonumber\\
& &
\end{eqnarray}
where $W=\hbar/\beta$ and
\begin{eqnarray}
Q_{0}(\beta)&=&\frac{1}{1-e^{-\beta}}\nonumber\\
Q_{n}&\equiv& Q_{0}^{-1}\frac{d^{n}}{d\beta^{n}}Q_{0}(\beta)\nonumber\\
Q_{1}&=&1-Q_{0}\nonumber\\
Q_{2}&=&1-3Q_{0}+2Q_{0}^{2}\nonumber\\
Q_{3}&=&1-7Q_{0}+12Q_{0}^{2}-6Q_{0}^{3}
\end{eqnarray}
Accordingly, we have
\begin{equation}
Q(\beta,\hbar)\approx Q_{0}(\beta)(1-W^{2}\beta^{3}Q_{0}(Q_{0}-1)^{2})
\end{equation}
Consequently, the average energy E and the specific heat C are
 respectively given  by
\begin{eqnarray}
E&=&-\frac{Q'}{Q}=
E_{0}+\hbar^{2}\frac{d}{d\beta}(\beta Q_{0}(Q_{0}-1)^{2})\nonumber\\
C&=&C_{0}-\beta^{2}\hbar^{2}\frac{d^{2}}{d\beta^{2}}(\beta Q_{0}(Q_{0}-1)^{2})
\end{eqnarray}

It is worth discussing the behavior of the above expression of $Q$ for given
$W$ and $\beta\rightarrow 0$. Since for small $\beta$ $Q_{0}\approx 1/\beta$,
then
\begin{eqnarray}
Q(\beta)&\approx& Q_{0}(1-W^{2} + W^{4}+...)\nonumber\\
E(\beta)&\approx&E_{0}-2W^{2}\nonumber\\
C(\beta)&\approx&C_{0}-6W^{2}
\end{eqnarray}
Accordingly, the above representation of $Q$ is valid only for
$\beta^{2}>6\hbar^{2}$, that is for $W\leq 1/\sqrt{6}$.

\bigskip

\subsection{\bf Partition function for given $\beta/\hbar$}

\bigskip

For given $\zeta=\beta/\hbar$ we can rely on the following inequalities for the
partition function
\begin{equation}
\int_{0}^{\infty}e^{-\beta E_{q}(x)}dx<Q(\beta,\hbar)<1+\int_{0}^{\infty}e^{-
\beta E_{q}(x)}dx
\end{equation}
where  $E_{q}(x)$ is a monotone function of $x$ which coincides with $n_{q}$
for
$x=n$.
 The above integral, which represents the classical limit for the partition
 function, can be represented as a combination of Weber function $E_{0}$ and
Neumann function $N_{0}$ [18],
\begin{equation}
\int_{0}^{\infty}e^{-\beta E_{q}(x)}dx=-\frac{\pi}{4\hbar}[N_{0}(\zeta
)+E_{0}(\zeta)]
\end{equation}
being $\zeta=\beta/\sinh~\hbar$.
 In particular,
\begin{equation}
E_{0}(\zeta)=-\sum_{n=0}^{\infty}\frac{(-1)^{n}(\zeta/2)^{2n+1}}{\Gamma^{2}
(n+3/2)}
\end{equation}

It is noteworthy that for fixed $\zeta$ there exists a limiting value $\bar{h}$
 of $\hbar$ such that for $\hbar<\bar{h}$ the partition function is so large
to coincide with the classical one. In fact, for $\zeta$ very small we have
\begin{equation}
-\frac{\pi}{4\hbar}[E_{0}(\zeta)+N_{0}(\zeta)]\approx \frac{1}{2\Gamma^{2}(1/2)
\zeta}+O(\zeta^{-2})
\end{equation}
Hence, for $\zeta<.1$ the classical and quantum partition
 functions are almost coincident for $\zeta>10 \hbar$.

In conclusion, while the quantum partition function of the q-oscillator is well
approximated by the classical limit for $\hbar$ going to zero, this is not
true for the classical harmonic oscillator.

Next, we consider the average energy $E$ for which we can write the following
inequalities
\begin{equation}
E_{<}\leq E\leq E_{>}
\end{equation}
where the lower ($E_{<}$) and upper ($E_{>}$) bounds are given by
\begin{eqnarray}
E_{<}&=&\frac{-\frac{1}{e\beta}+\int_{0}^{\infty}E_{q}(x)e^{-\beta E_{q}(x)}dx}
{1+\int_{0}^{\infty}e^{-\beta E_{q}(x)}dx}\nonumber\\
E_{>}&=&\frac{\frac{1}{e\beta}+\int_{0}^{\infty}E_{q}(x)e^{-\beta E_{q}(x)}dx}
{\int_{0}^{\infty}e^{-\beta E_{q}(x)}dx}
\end{eqnarray}

 Plugging (6.10) into the above expressions  yields
\begin{equation}
\frac{(2\sinh\hbar W)/(\pi e)-E_{0}^{'}(\zeta)-N^{'}_{0}(\zeta)}
{-4\hbar/{\pi} + E_{0}(\zeta)+N_{0}(\zeta)
}<E \sinh~\hbar<\frac{(2 \sinh\hbar W)/(\pi e) -
E_{0}^{'}(\zeta)-N^{'}_{0}(\zeta)}
{E_{0}(\zeta)+N_{0}(\zeta)}
\end{equation}
When $\hbar$ is so small to satisfy the following inequalities
\begin{eqnarray}
 \frac{2\hbar}{\pi}\ll E_{0}(\bar{W}^{-1})+N_{0}(\bar{W}^{-1})\nonumber\\
\frac{2\sinh~\hbar\bar{W}}
{\pi e}\ll -E^{'}_{0}(\bar{W}^{-1})-N^{'}_{0}(\bar{W}^{-1})
\end{eqnarray}
we can approximate $E$ by
\begin{equation}
E=-\frac{1}{\sinh~\hbar}\, \frac{E'_{0}+N'_{0}}{E_{0}+N_{0}}
\end{equation}
 Next, letting $\beta\rightarrow 0$ into the above equation we have
\begin{equation}
\beta E\approx-\frac{1}{\ln~\beta}
\end{equation}
Consequently, the specific heat decreases for $T\rightarrow\infty$ as
\begin{equation}
C\propto \frac{1}{\ln~T}
\end{equation}
 Thus the behavior of the specific heat of the q-oscillator is
 different from the behavior of the usual oscillator
in the high temperature limit. This property may serve for an experimental
check
of the existence of vibrational non linearity of the q-oscillator
fields.

\bigskip

\subsection{\bf Time-correlation function}

\bigskip

For characterizing adequately the thermalized q-oscillator it is worth
discussing the relative second-order correlation function $\gamma_{q}
(\beta,\hbar;t)$.
Defining
\beq
\tilde {Q}( \beta,\hbar;t) = \sum_{n=0}^{\infty} exp (-\beta
f^{2}(n) + it F(n-1)) ,
\eeq
we have
\beq
\gamma_{q}(z,\hbar;t) \dot{=} Tr (\rho a^{\dag}_{q}(t)a_{q}(0)) =- \frac {1}
{Q(\beta,\hbar)}\frac {\partial \tilde {Q}(\beta,\hbar;t)}{\partial \beta} .
\eeq

To compute $\gamma_{q}(\beta, \hbar;t)$ for $\hbar\rightarrow 0$
we can use the approximate expression of $\tilde {Q}(\beta,\hbar;t)$
\beq
\tilde {Q}(\beta,\hbar;t) \approx e^{it}\left[Q_{0}(\beta) - \hbar^2
\left(\beta \frac {<n^3> -<n>}{6} + it \frac{<n>-<n^2>}{2}\right)\right]~~,
\eeq
where
\beq
<n^k> = \sum_{n=0}^{\infty} n^{k} e^{-\beta n}~~~~k= 0,1,....
\eeq

It is worthwhile noting that $\gamma_{q}$ depends on the hierarchy of
photon distribution momenta $<n>, <n^{2}>, ...$. This circumstance marks the
main difference between q-oscillators and standard ones.
In fact, the dependence of the correlation function on the intensity is a
general property of q-oscillators, holding true also when the q-oscillator
is in a generalized coherent state. This property could
be used for testing experimentally the possibility that electromagnetic fields
behave like q-oscillators. In fact, accurate measurements of the dependence
of the field correlations (second and higher orders) on the field intensity
could help in finding upper bounds for the small quantity $\hbar$.

For $t=0$ the correlation function gives the mean value of $\hat{n}_q$,i.e.
\beq
\bar n_q = \gamma_{q}(\beta,\hbar;0).
\eeq
Such a distribution function is expressed in terms of Eq. (6.21) in which
$\tilde {Q}(\beta,\hbar;t)$ is taken at t=0.

 q-deformed Bose
distribution $\bar n$  can be obtained by the same method starting from the
Hamiltonian operator (2.23) with $\omega = 1$ and not from $H =a^{\dag}_q
a_q$ and one obtains
\beq
\bar n = \bar {n}_0 - \frac{\hbar^2}{6}\left[\frac{5}{2}((\bar {n^2})_0 -
 (\bar n)^{2}_0) + \frac{3}{2}((\bar {n^3})_0 - \bar {n}_{0}(\bar {n^2})_0)
+ \frac{3}{2}((\bar {n^4})_0 - \bar {n}_{0}(\bar {n^3})_0)\right] ,
\eeq
in which  $\bar{n}_0$
is the usual Bose distribution function and
\beq
\bar {n^k}_0 = 2\sinh \frac{\beta}{2} \sum_{n=0}^{\infty}
 n^{k} e^{-\beta (n+1/2)}~~~.
\eeq

\setcounter{chapter}{7}
\bigskip

\noindent{\bf 7. Conclusions}
\bigskip

We conclude this paper by pointing out some ideas of the work.\\ As we
understand now
the one-dimensional q-oscillator is nothing else than a nonlinear oscillator
with a very specific type of the nonlinearity. Namely,its frequency depends
on its energy as hyperbolic cosine of the energy. Thus, classical motion of
such
nonlinear oscillator is described as motion of q-oscillator. So, in this case
the frequency of the vibrations (or the velocity of motion) increases
exponentially with the energy. After standard quantization we obtain quantum
q-oscillator which is nothing else than a nonlinear quantum oscillator with
specific anharmonicity described by the infinite power series in energy.From
the discussion above it is now clear that there are different approaches to
generalyse this picture.In the frame of nonlinearities, even when the frequency
of vibrations depends only on the energy we could choose other functional
dependencies introducing functions different from hyperbolic cosine.
We could call these oscillators f-oscillators where $f$ is now the function
 determining the dependence of the frequency on the energy.This function
is related to the Poisson brackets of new $\alpha_f$-coordinates where
$\alpha_f=\alpha_f(\alpha\alpha^*)$ and the $\alpha$ is the harmonic oscillator
coordinate.Of course, this function has to contain a dependence on the
parameter
$\hbar$ such that for $\hbar\rightarrow 0$ the function $f\rightarrow 1$.\\
The quantum $f$-oscillator  obtained by canonical
quantization of the above is again a nonlinear oscillator with a
different dependence of $\omega$ on the energy. We could
also introduce other nonlinearities ny making the frequency to depend on
other constants of the motion, different from the energy.
 This gives the possibility to undertake a classification
 of such nonlinearities
and especially it is worthy for multidimensional oscillator.Here we may
introduce a deformed nonlinear motion considering the frequencies of the
oscillators depending either on their own energies or  on energies of some
groups of other oscillators. So, to preserve the U(N)-symmetry of
multidimensional isotropic oscillator we may introduce the frequency dependence
on energy
to be a function of total energy of all the oscillators. If each oscillator is
deformed as independent one we destroy the U(N)-symmetry of the initially
isotropic multidimensional oscillator.

In the quest for  adequate field theories, we limited ourselves to consider
some possible effects in systems which can be represented as a collection of
single oscillators. In particular,  we showed how the specific heat formula is
changed by the q-non-linearity. In addition, the second-order correlation
function depends on the average occupation number of the q-oscillator. This
property marks the main difference with the standard oscillator.

In spite of mathematical beauty of q-deformation procedure, (and
$f$-deformation too) the most important issue is to envisage  possible
experimental tests of the possibility of describing electrodynamic phenomena by
means of these q-oscillators. At a first thought , non-linear optics
experiments appear as potential candidates for obtaining upper bounds for the
value of $\hbar$. However, these experiments require the development  of a
q-field formulation of QED. For example, if a laser
beam behaves like a collection of q-oscillators interacting with the
matter in the standard way, an optical rectification experiment
carried out at different laser intensities could be used for testing the
deviation of the boson spectrum from the ideal equispaced pattern.
As an instance, apparatus
presently developed for carrying out accurate interferometric experiments
(f.i. LIGO in USA and VIRGO in Italy-France) on gravitational waves,
in view of their ability to reducing
the photon noise limit by averaging over periods of the order of one year
could in future become ideal candidates for performing these tests.

Another class of experiments could be based on the dependence of
time-correlation function of q-oscillator on the intensity. The interpretation
of these experiments does not depend critically on the development of a
q-field QED. In fact, these measurements can be carried out in vacuum. For
example, a continuous laser beam could be split in two beams of
different intensities  and sent to two spectra analyzers. Measuring the
widths of the two spectra it would be possible to put an upper
limit to the deformation parameter
$\hbar$. These measurements could last for years thus guaranteing a
signal-to-noise
ratio adequate for appreciating very small values of the parameter.

 The
non-linearity may also produce sqeezing [19] or may be related to non locality
and lattice structure in space and time. The q-oscillator and squeezing has
been discussed in [20]. The non locality in time has been studied in
[21]
where a q-deformed linear version of Klein-Gordon equation has been suggested.
We would like to point at the recent work [22] where
similar ideas
are discussed. The connection of q-oscillator with anharmonic oscillator
has been also discussed in [23]. We will present a more detailed analysis of
possible generalizations and experimental consequences in other future
publications.

\bigskip

Acknowledgements.
One of us (I.V.M.) thanks INFN and University of Napoli "Federico II"
for the hospitality.
The authors thank L.Biedenharn and G.Dell'Antonio for fruitful discussions.

\newpage
\begin{center}
{\bf References}
\end{center}

\bigskip

\noindent
1. V.G. Drinfield, Quantum Groups, {\sl Proc. Int. Conf.\ of Math.},
MSRI Berkeley CA (1986), p. 798.

\noindent
2. P. Kulish, N. Reshetikhin, {\sl J. Sov. Math.} {\bf 23}, 2435 (1983).

\noindent
3. S. Woronowicz, {\sl Comm. Math. Phys.} {\bf 111}, 615 (1987).

\noindent
4. M. Jimbo, {\sl Int. J. Mod. Phys.} {\bf A4}, 3759 (1989).

\noindent
5. Y. Manin, {\sl Comm. Math. Phys.} {\bf 123}, 163 (1989).

\noindent
6. L. C. Biedenharn, {\sl J. Phys.} {\bf A22}, L873 (1989).

\noindent
7. A. J. Macfarlane, {\sl J. Phys.} {\bf A22}, 4581 (1989).

\noindent
8. R. M. Mir-Kasimov, Proc. of the 18th Group Theoretical Methods in Physics,
Colloquium, Moscow June (1990) in Lecture Notes in Physics, v. 382,
p. 215, Springer (1991).

\noindent
9. M. Chaichian, D. Ellinas, P. Kulish, {\sl Phys. Rev. Lett.} {\bf 65}
(980) (1990).

\noindent
10. M. Born, L. Infeld, {\sl Proc. Roy. Soc.} {\bf A143}(410)(1934) and
{\bf A147}(522)(1934).

\noindent
11. A. P. Polychronakos, University of Florida preprint HEP-89-23, (1989)

\noindent
12. R.Floreanini, V. P. Spiridonov and L. Vinet in Group Theoretical
    Methods in Physics, Proceedings of Moscow Colloquium 4-9 June, 1990,
    eds.V.Dodonov and V.Man'ko, Lecture Notes in Physics vol.382,
    Springer Verlag (1991).

\noindent
13. P.P.Kulish, ibid. p. 195.

\noindent
14. I.A. Malkin, V. I. Man'ko, {\sl Phys. Lett.} {\bf 32A}, 243 (1970).

\noindent
15. I.A. Malkin, V.I.Man'ko, Dynamical symmetries and coherent states
    of quantum systems, Nauka Publishers, Moscow (1979) (in Russian).

\noindent
16. V. I. Man'ko, G.Marmo {\sl Mod. Phys. Lett.}{\bf A7}, 3411 (1992).

\noindent
17. T.D.Lee, Particle Physics and Introduction to Field Theory, Harwood
Academic Publishers, Chur (1981).

\noindent
18. I.S.Gradshteyn and I.M.Ryzhik, Tables of Integrals, Series, and Products,
Academic Press, New York (1980).

\noindent
19. J. N. Hollenhorst, {\sl Phys. Rev.} {\bf 190}, 1669 (1979).

\noindent
20. E. Celeghini, M. Rasetti, G. Vitiello, preprint ``On squeezing and quantum
groups" University of Florence, 1991.

\noindent
21. J.Lukierski, A. Nowicki and H.Ruegg,  {\sl Phys.Lett.} {\bf293B}, 344
 (1992).

\noindent
22. S.V.Shabanov, Quantum and Classical Mechanics and q-deformed Systems,
BUTP 92/24 August 1992

\noindent
23. The statement that the physical nature of the quantum q-oscillator
    depends on its
    anharmonicity is due to M. Artoni (talk at the Workshop on Harmonic
Oscillators,
    Maryland, March 1992).

\end{document}